\documentclass[10pt]{article}

\usepackage{amsmath}
\usepackage{amssymb}

\usepackage{graphicx}

\usepackage{cite}

\usepackage{color} 

\usepackage[labelfont=bf,labelsep=period,justification=raggedright]{caption}

\makeatletter
\renewcommand{\@biblabel}[1]{\quad#1.}
\makeatother

\date{}

\begin{document}

\begin{flushleft}
{\Large
\textbf{Maximum likelihood evidence for Neandertal admixture in Eurasian populations from three genomes}
}
\\
Konrad Lohse$^{1}$, 
Laurent A. F. Frantz$^{2}$
\\
\bf{1} Institute of Evolutionary Biology, University of Edinburgh, UK
\\
\bf{2} Animal Breeding and Genomics Group, Wageningen University, The Netherlands
\\

$\ast$ E-mail: konrad.lohse@ed.ac.uk
\end{flushleft}

\section*{Abstract}
Although there has been much interest in estimating divergence and admixture from genomic data, it has proven difficult to distinguish gene flow after divergence from alternative histories involving structure in the ancestral population. The lack of a formal test to distinguish these scenarios has sparked recent controversy about the possibility of interbreeding between Neandertals and modern humans in Eurasia. We derive the probability of mutational configurations in non-recombining sequence blocks under alternative histories of divergence with admixture and ancestral structure. Dividing the genome into short blocks makes it possible to compute maximum likelihood estimates of parameters under both models. We apply this method to triplets of human Neandertal genomes and quantify the relative support for models of long-term population structure in the ancestral African popuation and admixture from Neandertals into Eurasian populations after their expansion out of Africa. Our analysis allows us -- for the first time -- to formally reject a history of ancestral population structure and instead reveals strong support for admixture from Neandertals into Eurasian populations at a higher rate ($3.4 \% - 7.9\%$) than suggested previously.

\section*{Author Summary}

The excess sharing of genetic variation between Neandertals and non-African populations has been interpreted as a signature of interbreeding between archaic Hominins, and modern humans. However, this pattern could also be explained by ancestral sub-structure in Africa, that pre-dates the Human-Neandertal divergence. We develop a likelihood method, that uses the information contained in the mosaic of genealogical histories in the genome to statistically distinguish between these two alternative scenarios of human history and estimate all relevant parameters from three genomes. Applying this method to analyse genome-wide variation in humans allows us to reject ancestral population structure in Africa in favour of Neandertal admixture into Eurasian populations. The new method should be particularly useful for extracting historical signal from ancient or rare samples.

\section*{Introduction}

Whole genome sequence data have made it feasible to detect low levels of ancestral admixture between recently diverged populations and species even from few individuals. An increasing number of genome-wide analyses are uncovering signatures of introgression between sister species in a large range of taxa \cite{Cui:2013, Eaton:2013, Lawniczak:2010, Kulathinal:2009, Mallet:2012} suggesting that reticulations may be an ubiquitous feature of speciation. Similar evidence for gene flow after divergence has been found in Hominid lineages \cite{Patterson:2006}. A number of recent studies analyzing the Neandertal genome have suggested that admixture also occurred in the genus \emph{Homo} (i.e.\ from Neandertals and other archaic lineages into modern Eurasian populations) following the expansion of modern Humans out of Africa \cite{Green:2010, Yang:2012, Sankararaman:2012}. 

\begin{figure}[h]
\caption{Models of divergence between three populations with either A) a recent instantaneous, unidirectional admixture event (IUA model) or B) persistent structure in the ancestral population (AS model). Both histories lead to an excess of incongruent genealogies with topology ((a,b),b) (shown in blue) but different branch length distributions (Fig.\ S1).}
\includegraphics[totalheight=5in]{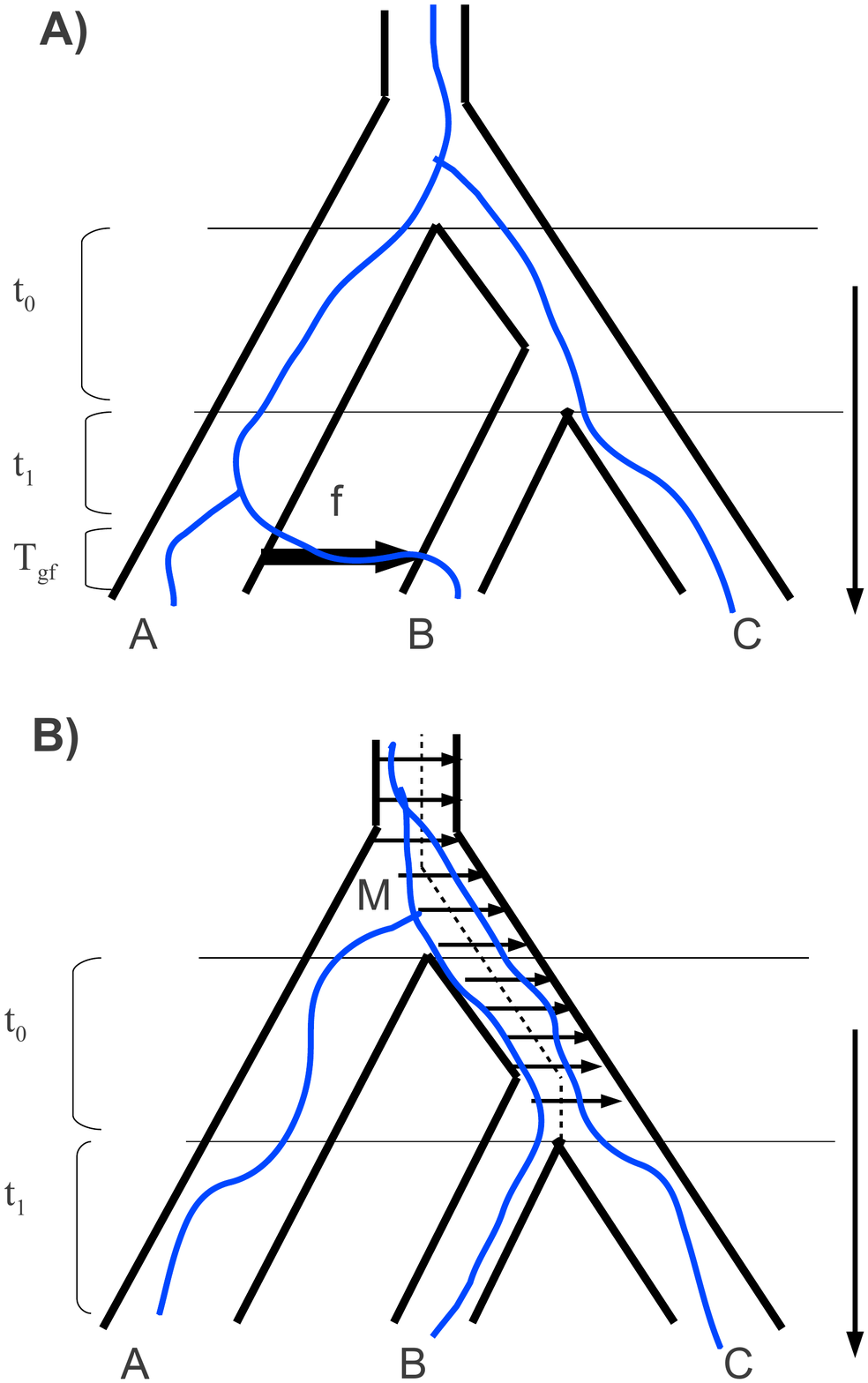}
\label{F:1}
\end{figure}

To test for admixture between Neanderthal and Eurasian populations, Green et al.\ \cite{Durand:2011, Green:2010} have developed a simple summary statistic. The D-statistic assesses the fit of a strictly bifurcating species tree. For a triplet of African, Eurasian and Neandertal genomes, and an outgroup (Chimpanzee), in which the underlying species tree is (African,  Eurasian, (Neandertal)), incomplete lineage sorting leads to two diagnostic site patterns. Denoting the ancestral state at a polymorphic site as $A$ and the derived state as $B$, mutations incongruent with the species tree may either be "ABBA" (i.e.\ shared by Eurasian and Neandertal) or "BABA" (shared by African and Neandertal). Given the inherent symmetry of coalescence in the common ancestral population under a null model of strict divergence without gene flow, the ratio $D=(N_{ABBA}-N_{BABA})/(N_{ABBA}+N_{BABA})$ is not expected to be significantly different from 0 \cite{Durand:2011, Green:2010} . In contrast, an excess of either ABBA or BABA sites cannot be explained by incomplete lineage sorting, suggesting population structure or gene flow (fig.\ \ref{F:1}).

Positive $D$, which may be indicative of gene flow, has been reported not only in the Neandertal analysis \cite{Green:2010}, but also in genome wide studies of closely related species of \emph{Heliconius} butterflies (whose origin is thought to have involved the introgression of color pattern genes \cite{Mallet:2012}) and an island radiation of pigs in South East Asia \cite{Frantz:2013}. 

$D$ is a drastic summary of genetic variation and -- like other population genetic summary statistics such as $F_{ST}$ -- suffers from the fundamental limitation that it is not diagnostic of any specific historical scenario. In particular, Durand et al.\ \cite{Durand:2011} contrast the expectation of $D$ under a model of instantaneous unidirectional admixture (IUA) (fig.\ \ref{F:1}A) and a different divergence model, involving structure in the ancestral population (AS) (fig.\ \ref{F:1}B). The AS model assumes a genetic barrier (with gene flow of $M=4N_e m$ migrants per generation) which arises in the common ancestral populations and persists until the most recent split \cite{Durand:2011}. Under this model, increasing barrier strength leads to increasing topological asymmetries \cite{Slatkin:2008} and hence positive $D$. A key finding of the Durand et al.\ \cite{Durand:2011} analysis is that it is impossible to distinguish between gene flow after divergence and structure in the ancestral population using $D$. Although Green et al.\ \cite{Green:2010} argue that admixture from Neandertals into Eurasians is the most plausible history, they conclude that "we cannot currently rule out a scenario in which the ancestral population of present-day non-Africans was more closely related to Neandertals than the ancestral population of present-day Africans due to ancient substructure within Africa." This has lead to recent controversy about the genomic signature of Neandertal admixture. In particular, Manika and Eriksson \cite{Manica:2012} have used Approximate Bayesian Computation to show that $D$ values identical to those observed in the human-Neandertal triplets can be generated under stepping-stone type models of colonization and structure without admixture and "[...] recommend caution in inferring admixture from geographic patterns of shared polymorphisms". In contrast, recent studies examining patterns of linkage disequilibrium \cite{Sankararaman:2012} and allele frequency spectra of modern human populations \cite{Yang:2012} provide qualitative support for Neandertal admixture. However, a rigorous statistical comparison of these alternative scenarios of human history is lacking.

Importantly, $D$ exclusively uses information contained in the mean length and frequency of genealogical branches. However, given the randomness of the coalescent process, much information about population history is contained in the higher moments of the distribution of branch lengths. An obvious strategy for exploiting this information is to partition the genome into short sequence blocks within which recombination can be ignored, and to maximize the joint likelihood across blocks \cite{Nielsen:2001,Yang:2002,Zhu:2012} .

In this study we develop a method to compute maximum likelihood estimates of parameters under the AS and IUA models from genomic triplets. Assuming an infinite sites mutation model and an outgroup to polarize mutations, the information in a block of sequence can be summarized by counting the number of mutations on each genealogical branch. Lohse et al. \cite{Lohse:2011b} show that for an arbitrary model of history the probability of a particular mutational configuration can be calculated from the Generating function (GF) or Laplace Transform of the distribution of genealogical branch lengths which has a simple, recursive form. The GF is derived for the IUA and AS models in the Methods. Assuming for now that sequence blocks are not affected by linkage, the logarithm of the likelihood ($lnL$) for a given multilocus dataset is simply the sum of $lnL$ across blocks. Because the number of mutational configurations is limited and the $lnL$ for each configuration only needs to be tabulated once, this numerical computation can deal with an arbitrary number of blocks and is far more efficient than simulation-based methods \cite{Gronau:2011, Hammer:2011}.

Below we first contrast the power of the new method with that of the $D$ statistic and then apply it to triplet samples of contemporary human genomes from Africa and Eurasia and the Neandertal genome \cite{Green:2010,Yang:2012,Wall:2013} to quantify the relative support for the AS and IUA model. Finally, we use simulations to demonstrate the robustness of our inference to the effect of recombination. 
 
\section*{Results}

\subsection*{Power analyses}
\begin{figure}[h]
\caption{A) The expected difference in support ($E[\Delta lnL]$) between A) the IUA model and the AS model (bold) and between the IUA and a null model of strict divergence (dashed), when IUA is true plotted against the admixture fraction $f$. B) shows analogous results for $E[\Delta lnL]$ against barrier strength ($1/M$) when the AS model is true. Plots are based on analytic results for the likelihood and assuming 10,000 sequence block, $\theta=3$ and the time parameters of Durand et al.\ \cite{Durand:2011} (Table S1).}
\includegraphics[totalheight=2in]{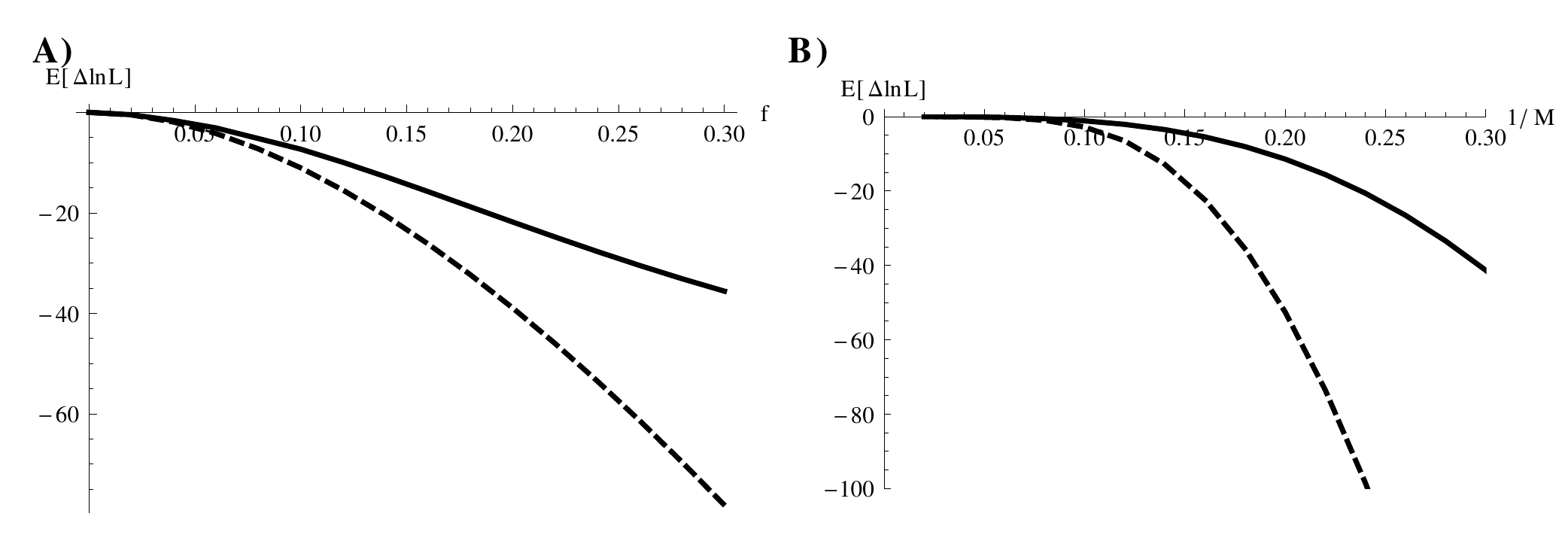}
\label{F:2}
\end{figure}

We investigated the power of the likelihood method analytically and compared it to that of the $D$ statistic. For ease of comparison, we focused on the history previously studied by Durand et al. \ \cite{Durand:2011} (Table S1). Our analysis highlights several advantages of the maximum likelihood scheme:

Firstly and as shown in figure \ref{F:2}, the likelihood method can distinguish between ancestral admixture (IUA) and ancestral structure (AS) models regardless of which scenario is true. 

Secondly, there is greater power (as measured by $E[\Delta lnL]$) to distinguish between the IUA history (when true) and a null model of strict divergence using maximum likelihood estimation on 10,000 unlinked sequence blocks compared to $D$ calculated from same number of unlinked SNPs. This is true even if we set the length of blocks such that they contain a single SNP on average (fig.\ S2A).

Finally, we can use Fisher Information a measure of the sharpness of the likelihood surface (see Methods) to quantify how informative sequence data are about a particular model parameter, and hence how accurate one can expect parameter estimates to be. For example, under the IUA history, there is much more information about the admixture fraction $f$ than the time of admixture $T_{gf}$ (Table S1). E.g.\ given a sample of 10,000 blocks of 2kb length, one would expect a standard deviation (SD) of 0.0145 for estimates of $f$ but 0.178 for $T_{gf}$ (Table S1). Note that in contrast to the $D$ statistics which have been used to derive a lower bound on $f$, the maximum likelihood estimate of $f$ is unbiased \cite{Durand:2011}.

As expected, increasing the length of sequence blocks, sharpens the likelihood surface (fig.\ S2) and so increases both the power of the likelihood method to distinguish alternative models (fig.\ S2A) and the accuracy of parameter estimates (Table S1, fig.\ S2B).


 
\subsection*{Application to human-Neandertal data}

We consider the Neandertal genome \cite{Green:2010} and three high-quality individual human genomes: Yoruba (YRI), French (CEU) and Han (CHB) obtained from complete genomics (Methods). Following Green et al.\ \cite{Green:2010}, we excluded all transition substitutions as these are more prone to ancient DNA damage \cite{Briggs:2007} and only used autosomal chromosome sequence. We focused our analysis on two triplet combinations, Neandertal/Eurasian/Yoruba, where the Eurasian genome is either CEU or CHB. Sites were polarized (ancestral vs.\ derived) using the sequence reconstruction of the Human-Chimp ancestor. We divided the human genome into blocks of fixed length after filtering (Methods). Our initial block length of 2kb of covered sequence yielded a total of 146,281 blocks, each with an average of 1.85 mutations in the ingroup triplet.

We computed maximum likelihood estimates of parameters under the IUA model (with one or two ancestral $N_e$ parameters), the AS model and a null model of strict divergence. The effect of physical linkage between blocks can be ignored when computing point estimates of parameters which are unbiased regardless. However, in order to obtain confidence intervals and compare the relative support between models, we need to remove the effect of genetic linkage between blocks. In our likelihood framework, this can be done by rescaling estimates obtained from the full data (Methods). To be conservative, we assumed that statistical association due to genetic linkage are negligible at distances >100kb \cite{Sankararaman:2012}.

The IUA model provided a much better fit to the data than both a null model without gene flow and the AS model (Table \ref{T:1}). Note that the differences in support ($\Delta lnL$) between the null and the IUA model are highly significant assuming a $\chi^2$ distribution, which is conservative. Allowing the size of the ancestral population between the two divergence times to differ from that of the common ancestral population (the IUA$_2$ model) further improved model fit, although the $lnL$ improvement was marginal for 2kb blocks (but see next section).

To convert estimated divergence times (scaled in $2 N_e$ generations) into absolute values, we followed Green et al.\ \cite{Green:2010} and assumed an average gene divergence time between chimps and humans of 6.5 MY and a generation time of 25 years. Given this calibration, we estimated that Neandertals diverged from the ancestor of modern humans 329--349 KYA ($T_2$). The divergence between African and non-African human populations, i.e. the second "Out-of-Africa" event ($T_1$) occurred 122--141 KYA. Estimates for $T_1$ and $T_2$ generally agreed well between the CEU and CHB analyses (Table \ref{T:2}, Table S2). We inferred a fraction of Neandertal admixture ($f$) of 5.9 and 5.3 \% for the CHB and CEU respectively with 95 \% C.I.\ broadly overlapping between the two analyses  (fig.\ S3). There was very little information about the time of admixture and the 95 \% C.I. for this parameter included $T_1$ in all analyses (Table \ref{T:2}, S2).

\subsection*{Sensitivity analyses}

In practice, the assumption that mutations in the same sequence block are completely linked limits multilocus analyses to relatively short blocks. Because of this, the usefulness of our method depends on the relative rates of recombination and mutation and the heterogeneity of both processes along the genome. There is a clear trade off between power and bias: if blocks are too short, they contain little additional information compared to SNP frequency spectra. Choosing excessively long blocks on the other hand potentially biases parameter estimates because recombination within blocks reduces the variance in inferred branch lengths \cite{Hudson:1985} and blocks with detectable recombination breakpoints (4-gamete criterion) need to be excluded. We investigated the influence of intra-locus recombination on parameter estimates in two ways.

Firstly, we repeated all analyses with longer blocks (4kb and 8kb). Increasing block length did not change the relative support for alternative models (Table \ref{T:1}). However, as expected from the analytic results (Table S1 and fig.\ S2), using longer blocks increased power (Table \ref{T:1}). For example, in the 4 and 8kb datasets one would be able to accept the more complex IUA$_2$ model with two ancestral $N_e$ parameters. Although in general, inference was little affected by block length (Table \ref{T:1} and S2 and fig.\ S3), we observed subtle shifts in parameter estimates. Estimates of divergence and admixture times increased, whereas the inferred ancestral $N_e$ decreased with block length. In contrast, the $N_e$ between $T_1$ and $T_2$ (in the IUA$_2$ model) increased with block length (Table S2). Secondly, we applied the maximum likelihood computation to data simulated with recombination. This confirmed that -- assuming a genome wide recombination rate of 1.3 cM/Mb \cite{Li:2011} and 2-8kb blocks --  the expected biases in estimates of divergence time and $f$ are negligible (fig.\ S4).

Our analysis ignores mutational heterogeneity across loci. To test whether this could affect inference, we partitioned 2kb blocks into 10 bins of equal size according to their relative distance to the chimpanzee. Perhaps surprisingly, incorporating relative mutation rates for each bin resulted in lower support overall 
but little change in parameter estimates.

As a simple way to assess the overall fit of the data to the inferred history, we compared the observed distribution of the total number of mutations ($S$) in each topology class with its expectation. Table S3 shows a close match between observed and expected frequencies. The only notable disagreement is a slight overall excess of topologically resolved blocks (2 \%) and a subtle excess of blocks with an incongruent topology (e.g.\ (YRI,(N,CEU)) or (CEU,(N,YRI))) and a short genealogy as indicated by low $S$ (see $S=1$ in Table S3). This may be a result of selective constraints on some sequences, which are not captured by our method.

\section*{Discussion}

We have developed a method to numerically fit alternative models of divergence between three populations with either recent gene flow or ancient structure to genomic data. Partitioning the genome into short sequence blocks within which recombination can be ignored provides an efficient way to compute maximum likelihood estimates under these models. Both the agreement of parameter estimates across a range of block sizes (Table S2) and our sensitive tests on simulated data (fig.\ S4) highlight the robustness of this approach to intra-locus recombination. Clearly, treating nearby SNPs as linked over short distances is a realistic approximation which adds substantial information to historical inference.

Our maximum likelihood method has several advantages over the $D$ statistic \cite{Green:2010, Durand:2011}: First, provided the assumptions about recombination and mutation can be justified, it is statistically optimal in the sense that all available information is used and therefore has greater power. Second, instead of testing a null model, one obtains joint estimates of all relevant parameters under a set of alternative models. This constitutes a substantial improvement over previous genomic analyses that have estimated divergence and admixture parameters separately and using different approaches. Finally, and in contrast to the assertion of Durand et al.\ \cite{Durand:2011} that distinguishing between the ancestral admixture (IUA) and population structure (AS) "[...] will require using more than one sample per population", our analysis shows that the two scenarios can indeed be distinguished from minimal samples. Considering the difference in the length distribution of branches between these models  (fig.\ S1), it is clear where the signal comes from. While the length distribution of internal branches differs only subtly between the two models, there is a marked difference in the distribution of external branches: incongruent genealogies with short external branches (i.e.\ $t_{ex} < T_1$) are possible under the IUA model, but not the AS model (A vs.\ B in fig.\ S1).

\section*{Conclusions about Human history}

Our analysis of human-Neandertal data provides strong statistical support for the IUA model and confirms previous claims that Neandertals contributed genetically to contemporary Eurasian populations \cite{Green:2010, Yang:2012, Sankararaman:2012}. However, in contrast to previous studies we can conclusively reject long-term population structure in the ancestral African population as an alternative explanation for the excess sharing of derived mutations by Neandertal and Eurasians. 

The parameter estimates we infer agree well with a number of recent population genomic studies on human history \cite{Green:2010, Yang:2012, Sankararaman:2012, Wall:2013}. For example, our population divergence times match those of Green et al.\ \cite{Green:2010} and the ancestral population size is close to the average $N_e$ inferred by Li and Durbin \cite{Li:2011} during that period (120-500KY). Similarly, our inference of a slightly higher fraction of Neandertal admixture in the Han compared to the European genome (Tables \ref{T:2} and S2) mirrors recent findings based on comparing average $D$ in Asian and European individuals \cite{Wall:2013}. 

It is notable that we infer a larger fraction of Neandertal admixture ($3.4\% >f> 7.9\%$) than previous studies (1-6 \% \cite{Green:2010, Durand:2011}). This difference is to be expected given that the $D$-based estimator is a lower bound of $f$ \cite{Durand:2011}, while -- all else being equal -- maximum likelihood estimates are unbiased. While our exploration of simulated data show that ignoring recombination within blocks slightly biases $f$ estimates upwards and so leads to larger $f$ estimates for longer block (fig.\ S4), we observe little such bias in the Neandertal analysis (fig.\ S3 and S4). We also re-iterate the point made by Durand et al.\ \cite{Durand:2011} that $f$ estimates are rather sensitive to assumptions about the effective population sizes of Neandertals. We have followed Durand et al.\ \cite{Durand:2011} in assuming the $N_e$ of Neandertals to be equal to that of the common ancestral population. It will be interesting to incorporating information about the $N_e$ of Neandertals into such analyses in the future.

Although in principle, our method allows us to estimate the time of admixture $T_{gf}$ and our estimates for this parameter encompass those of Sankararaman et al.\ \cite{Sankararaman:2012} (37KY--86KY), our power analysis shows that multilocus data contain little information about this parameter (Table S1). This makes intuitive sense considering that only mutations that arise between $T_{gf}$ and $T_1$ contribute information about this parameter. Methods that use information contained in patterns of linkage \cite{Sankararaman:2012, Coop:2013} are more informative over such recent time scales.

In conclusion, we show that maximum likelihood calculations on blocks of sequences allow for a joint estimation of divergence times, ancestral effective population sizes and the fraction and time of admixture. This approach has greater power than summary statistics and can distinguish between subtly different scenarios of admixture and ancestral population structure. Our results allows us to conclusively reject the ancestral admixture model and demonstrate that secondary admixture from Neandertals into Eurasians took place after the expansion of modern humans out of Africa. This has important implications for our understanding of human evolution. Future studies, based on ancient and/or modern DNA will likely shed light on the frequency at which such reticulation events took place in the Hominin lineage. Because our approach maximizes the information contained in a single individual per taxon, it will be particularly useful for revealing the history of rare and extinct species and populations for which samples are limited. Another advantage of considering minimal samples is that it renders inferences of ancestral parameters robust to the details of more recent demographic events which would otherwise need to be modeled explicitly. Given that the analytic basis of our method is not restricted to any particular model \cite{Lohse:2011b}, it should be possible to use analogous calculations for other histories and incorporate recombination in these inferences explicitly in the future.

\section*{Materials and Methods}
\subsection*{Computing likelihoods}

We consider a model of divergence and admixture between three populations labeled $A$, $B$ and $C$, where $C$ is the older population and $B$ the population receiving migrants (fig.\ \ref{F:1}). Individuals sampled from these populations are labeled $a$, $b$ and $c$. Assuming an infinite sites mutation model and an outgroup information, the information in a block of sequence can be summarized as a vector of mutation counts $\underline{k}= \{k_a,k_b,k_c,k_{ab},k_{ac},k_{bc}\}$, where mutation types are labeled by the node in the genealogy they are connected to, i.e.\ $k_a$ is the number of mutations unique to sample $a$ and $k_{ab}$ the number of mutation shared by $a$ and $b$. We are interested in computing the probability of a particular mutational configuration at a block $P[\underline{k}_j]$. Lohse et al.\  \cite{Lohse:2011b} show that for an arbitrary model $P[\underline{k}_j]$ can be calculated by taking derivatives from the Generating function (GF) or Laplace Transform of the distribution of branch lengths $\underline{t}$ (analogous to the mutational counts $\underline{k}$). The GF of branch lengths is defined as $\psi[\underline{\omega]}=E[e^{-\underline{t}.\underline{\omega}}]$ and relates the sample configuration at a particular time in the ancestral process, $\Omega$ to the configuration $\Omega_i$ before some previous event $i$ \cite{Lohse:2011b}:

\begin{equation}
\psi [\Omega] = \frac{\sum _{i} \lambda _i \psi [\Omega _{i}]}
{\left(\sum _i \lambda _i+\sum _{|S|=1} \omega _S \right)}
\label{E:1}
\end{equation}

The denominator is given by the total rate of events $\sum _i \lambda _i$ plus the sum of dummy variables $\omega _S$ corresponding to branches involved in the event (for the first event these are the "leaves" of the genealogy, i.e.\ $|S|=1$). The GF under the IUA model is an extension of the GF for a model of strict divergence given by \cite{Lohse:2012}. For simplicity, we initially assume that both ancestral populations are of equal size. Following Lohse et al.\ \cite{Lohse:2011b}, we note that the above recursion for the GF only holds for a slightly different model in which the times in between discrete events (i.e.\ the time of admixture $T_{gf}$, $\tau_1$ and $\tau_2$, fig.\ \ref{F:1}A) are exponentially distributed random variables. We define corresponding time parameters measuring time from the present: $T_1 = T_{gf}+ \tau_{1}$ and $T_2 = T_{gf}+\tau_{1}+\tau_{2}$. The type of event that is possible in each interval is specified by the model: going backwards in time; we first only allow for an admixture event (with rate $\Lambda_{gf}$). During this event the lineage in population $B$ either traces back (instantaneously) to population $A$ (with probability $f$) or remains in population $B$ (with probability $1-f$). Once admixture has occurred, we allow for the merging of populations $B$ and $C$ (at rate $\Lambda_1$) and finally the merging of populations $A$ and the population ancestral to $B$ and $C$ (at rate $\Lambda_{2}$). The two population mergers correspond to divergence events forwards in time. Given the general recursion for the GF (eq.\ \ref{E:1}), we can write down the GF for each of the 12 possible sampling configurations in this model \cite{Durand:2011}. These are given in the online SI and are easily solved using \emph{Mathematica} \cite{Wolfram:2010} (see Supporting.nb).

We can recover the GF for the original model of discrete splits which we denote $P[\underline{\omega}]$ from $\psi[\underline{\omega}]$ by noting that $\psi[\underline{\omega}]=\int \Lambda_1\Lambda_2\Lambda_{gf} P[\underline{\omega}]e^{-\underline{\Lambda}.\underline{T}}d\underline{T}$. Thus multiplying $\psi[\underline{\omega}]$ by $(\Lambda_{gf} \Lambda_1 \Lambda_2)^{-1}$ and inverting once for each event with respect to the respective $\Lambda$ parameter gives the GF under the split model. Although this expression is cumbersome (see Supporting.nb), decomposing it into the contributions from the three different topologies \cite{Lohse:2011b} yields relatively compact formulae (online SI). Using equation \ref{E:1}, the GF for a model of ancestral structure (AS) can be derived analogously (Supporting.nb).

The probability of a particular mutational configuration at a locus $P[\underline{k}_j]$ can be calculated from $P[\underline{\omega}]$ by taking successive derivatives \cite{Lohse:2011b}. To calculate the likelihood for a given dataset, we tabulate the probabilities of all mutational configurations and take the product across  blocks. Code for this calculation is implemented in \emph{Mathematica} \cite{Wolfram:2010} (available from Dryad repository XXX). The sum of the logarithm of likelihoods across loci is maximized using the inbuilt \emph{Mathematica} function \emph{FindMaximum}. For a single dataset, this takes a few minutes on a modern desktop.

We can invert the GF to find the full distribution of branch length. Figure S1 shows these distributions for the internal branch ($t_{in}$) and the shorter external branches ($t_{ex}$) under both the IUA and AS models.

\subsection*{Power analyses}

We assumed the IUA history previously studied by Durand et al \cite{Durand:2011} to compare the likelihood method and $D$: $T_{gf}=2,500$, $T_1=3,000$, $T_2=12,000$ and $f=0.04$. Assuming $N_e=10,000$ (fixed for all populations) this roughly matches that previously inferred for Neandertals, African and Eurasian \emph{H.\ sapiens} by \cite{Green:2010}. All time parameters are in generations, corresponding values scaled in $2N_e$ generations are given in Table S1.

Given a dataset consisting of $j$ different mutational configurations $\underline{k}_i$ and a true history $H_T$, the expected difference in support, i.e.\ $E[\Delta lnL]$ for two alternative models $H_0$ and $H_1$ (one of which may be $H_T$) can be computed as:

\begin{equation}
E[\Delta lnL] = \sum_{i}^j (lnL[\hat \Theta_0|\underline{k}_i]-lnL[\hat \Theta_1 | \underline{k}_i]) \times P[\underline{k}_i|H_T]
\end{equation}

where $\hat \Theta$ denotes the set of parameter values that maximize $lnL$ under a particular model. Analogously, the accuracy of the likelihood method to estimate a particular model parameter $\theta$, can be quantified using Fisher information. This is defined as $I = -\frac{\partial^2 lnL}{\partial \theta ^2}$ and measures the sharpness of the $lnL$ curve near the maximum \cite{Edwards:1972}. The average information about a parameter contained in a sequence block is given by summing $I$ over all possible mutational configurations $j$ weighted by their probability:

\begin{equation}
E[I_i] = \sum_{i}^j -\frac{\partial^2 lnL[\hat \Theta |\underline{k}_i]}{\partial \theta ^2} \times P[\underline{k}_i|\hat \Theta]
\end{equation}

The expected information in a data set consisting of $n$ sequence blocks is simply $n \times E[I]$. Assuming parameter values are away from the boundaries, the inverse of $I$ gives a lower bound on the variance (and covariance) of parameter estimates \cite{Rao:1945}. 

\subsection*{Application to human-Neandertal data}

We downloaded BAM files (short-read alignment) of the three Vindija bones (SLVi33.16, SLVi33.25 and SLVi33.26) that were aligned to the human genome (hg18), from the UCSC genome browser (http://genome.
ucsc.edu/Neandertal). We only used sites with a minimum mapping quality of 90 and a sequence quality of 40 and filtered out sites that were covered by more than 3 reads, as the genome wide average depth of coverage was approximately 1.5 \cite{Green:2010}. We further excluded the first and last 5bp of every read, as these positions are enriched with sequecing errors \cite{Green:2010}. We also excluded transitions to limit the effect of ancient DNA damage \cite{Briggs:2007}. We obtained genotype files for a European (CEU; Coriell ID: NA06985), Han (CHB; Coriell: NA18526), and Yoruba (YRI; Coriell ID: NA18501) individual from the complete genomics website (ftp://ftp2.completegenomics.com, release 1.2). For the outgroup sequences, we extracted the genotype of the chimpanzee (\textit{Pan troglodytes}), and the Human-Chimp ancestor sequence reconstruction (available from the 4 primates EPO alignment provided by Ensembl release 54) in 1:1 human-chimp orthologous regions for each site that was covered in the Neandertal genome. Genotype files were filtered for transitions (on all branches) using custom perl scripts. We partitioned the human genome into 5, 10 and 20kb fixed length blocks. For each block, we sampled exactly the first 2, 4 or 8kb of sequence covered in all samples (three humans sequences, both outgroups and the Neandertal) and discarded any block with lower coverage.

Although the data were unphased, the low heterozygosity -- only 17 \% of SNPs were heterozygous in YRI, the most heterozygous individual -- and the short block lengths meant that the majority of sequence blocks contained no more than one heterozygous site per individual so that phase ambiguity within blocks was not an issue. Thus, for each site that was heterozygous in an individual (or in the case of the Neandertal a sample of three individuals), we simply chose one allele at random. Note that assuming an infinite site mutation model and a single genealogy underlying the polymorphisms in each block, heterozygous sites that are unique to one sample and invariable in all others can only arise due to mutations on external branches and so their phase does not affect the inferred topology (fig.\ S6).


While the analysis of Green et al.\ \cite{Green:2010} focuses on shared derived site, the likelihood method uses all site types. In fact, our analytic results show that much of the information to distinguish between the IUA and AS models is contained in the length distribution of external branches (fig.\ S1). This presents a problem: in the low coverage Neandertal sequence, it is challenging to distinguish true singletons from DNA degradation, sequencing and alignment error. To address this, we made a simple correction based on the symmetry of genealogies. Assuming that sequencing error in the modern human data can be ignored and that mutation rates and generation times are the same in Neandertals and modern humans, the proportion of true Neandertal singletons can be estimated from the difference in the number of divergent sites between humans and chimpanzee and between Neandertal and chimpanzee. We incorporated the estimated proportion of true Neandertal singletons (41 \%) by randomly sub-sampling derived sites unique to the Neandertal in each sequence block with this probability. Note that ignoring the fact that Neandertals died out, is consistent with both our model and this correction and so does not bias parameter estimates.

Violations of the 4-gamete criterion within a block can arise either due to recombination, back-mutation or phasing error, all of which are incompatible with our assumptions. We therefore excluded blocks containing more than one type of shared derived mutation from the analysis (1.5 \%, 4.9 \% and 14.2 \% in the 2, 4 and 8 kb datasets respectively). Applying the inter-block distance and filtering steps described above to the entire human autosome, yielded 291,620, 146,281 and 71,940 blocks of 2kb, 4kb and 8kb length respectively. Corresponding input files for \textit{Mathematica} and code for our maximum likelihood analyses are deposited on the Dryad repository (no XXX).

In order to remove the effect of physical linkage on our analyses, we assumed that LD between blocks separated by a distance of 100kb can be ignored. Thus, we rescaled $\Delta lnL$ between models by a factor of $(100kb/l)^{-1}$ and 95 \% C.\ I.\ of parameters by a factor $\sqrt{(100kb/l)}$, where $l$ is the physical length of blocks. Although, the 100kb threshold is arbitrary, the above can be used to adjust our results for any level of linkage.

To quantify the bias in parameter estimates due to intra-locus recombination, we simulated data under the best fitting model estimated from the 2kb CEU data (Table \ref{T:2}) for varying block lengths (1-8kb) and assuming a human recombination rate of 1.3 cM/Mb.

\section*{Acknowledgments}
We would like to thank Nick Barton for discussions and comments. Comments from Joshua Schraiber, Nick Patterson, Thomas Mailund and two anonymous reviewers on earlier versions of this manuscript greatly improved this work. We are also indebted to Lynsey McInnes for help with simulations. This study was supported by a fellowship from the UK Natural Environment Research Council to KL (NE/I020288/1) and funding from the European Research Council (249894) to LF.

\bibliography{thesisbiborg}
\clearpage

\section*{Tables}

\begin{table}[h]
\caption{Support $\Delta lnL$ relative to the best fitting model for alternative model of history: Strict divergence (Null), divergence with admixture (IUA) or ancestral population structure (AS). The IUA$_2$ allows for two different ancestral $N_e$.}

\label{T:1}
\begin{tabular}{llllll}
\hline
Dataset			&IUA$_2$ (5)		&IUA (4)	&AS (4)		&Null (3) \\
\hline	
CHB, 2kb		&0			&\textbf{0.25}	&9.47 	&9.47\\
CEU, 2kb		&0			&\textbf{0.15}	&9.15 	&9.15\\
\hline	
CHB, 4kb		&\textbf{0}			&5.70	&15.75 	&32.70\\
CEU, 4kb		&\textbf{0}			&6.47	&14.94 	&33.05\\
\hline	
CHB, 8kb		&\textbf{0}			&27.92	&37.77 	&86.96\\
CEU, 8kb		&\textbf{0}			&27.95	&34.95 	&82.35\\
\hline	
\end{tabular}
\end{table}

\begin{table}[h]
\caption{Maximum likelihood estimates of parameters under the divergence with admixture (IUA) model. Time parameters are scaled in $2 N_e$ generations and measured from the present. The second row (in bold) gives absolute parameter values, i.e.\ effective population sizes in individuals and divergence in KY. 95\% confidence intervals (in brackets) were calculated assumption that LD between block $>100kb$ apart can be ignored. Estimates obtained by Green et al.\ \cite{Green:2010} and Durand et al.\ \cite{Durand:2011} for comparison}
\label{T:2}
\begin{tabular}{lllllll}
\hline	
dataset			&$\theta$	&$T_1$			&$T_2$				&$T_{gf}$ 			&$f$ \\
\hline	
CHB, 2kb		&0.423		&0.376			&0.968 				&0.217				&0.059, (0.034--0.072) \\
 &\textbf{7,000, (6,950--7,190)}	&\textbf{132, (122--141)}	&\textbf{339, (329--349)}		&\textbf{75.8, (0--$T_1$)}	&~\\
CEU, 2kb		&0.423		&0.379			&0.967 				&0.157				&0.053, (0.039--0.079) \\
				&\textbf{7,012, (6,950--7,190)}	&\textbf{132, (123--142)}	&\textbf{339, (329--349)}		&\textbf{55.1, (0--$T_1$)} &~\\
\hline					
 &\textbf{10,000} &n/a	&\textbf{270--440KY}	 &n/a	& 0.01--0.06* \\
\hline	
\end{tabular}
\end{table}
\clearpage

\section*{Figures}

\end{document}


\title{Supplementary Information -- Maximum likelihood evidence for Neandertal admixture in Eurasian populations from three genomes}
\author{Konrad Lohse$^{1}$, Laurent, A. F. Frantz$^{2}$}
\date{}
\maketitle

\date{}

\maketitle 
\begin{center}

$^{1}$Institute of Evolutionary Biology\\
University of Edinburgh\\
Kings Buildings\\
Edinburgh EH9 3JT, UK

$^{2}$ Animal Breeding and Genomics Group\\
Wageningen University\\
Wageningen, The Netherlands

\end{center}

\newpage

\section*{GF derivation and likelihood calculations}

Under the IUA model, there are 12 sampling configurations and 12 corresponding GF equations in total. Since most of these are identical to those for the strict divergence model \citep[see][Appendix]{Lohse:2012}, we only give extra terms here. We denote the GF for the sampling configuration before the admixture event $\psi[*a/b/c]$ (where populations are separated by / ). Going backwards in time, the first event is admixture which occurs at rate $\Lambda_{gf}$: with probability $f$ the $b$ lineage traces back to population $A$ leading to configuration $a,b/\emptyset/c$ (where $\emptyset$ denotes an empty population), with probability $1-f$ the $b$ lineage remains in its population giving configuration $a/b/c$. The GF is:
\begin{equation}
\psi[*a/b/c]=\frac{\Lambda_{gf}}{(\Lambda_{gf}+\omega_a+\omega_b+\omega_c)}\left(f \psi[a,b/\emptyset/c]+(1-f)\psi[a/b/c] \right)
\end{equation} 

Movement of the $b$ lineage leads to a total of three subsequent configurations that are not possible without admixture: $a,b/\emptyset/c$ (both a and b lineages are in population A), $\{a,b\}/\emptyset/c$ ($a$ and $b$ have coalesced and are in population $A$ prior to $T_1$) and $\{a,b\}/c$ ($a$ and $b$ have coalesced and are in population $A$ after $T_1$). The corresponding GF terms are:

\begin{equation}
\begin{split}
\psi[a,b/\emptyset/c]&=\frac{1}{(1+\Lambda_1+\omega_a+\omega_b+\omega_c)}\left(\psi[\{a,b\}/\emptyset/c]+\Lambda_1 \psi[a,b/c] \right)\\
\psi[\{a,b\}/\emptyset/c]&=\frac{\Lambda_1 \psi[\{a,b\}/c]}{\Lambda_1+\omega_{ab}+\omega_c}\\
\psi[\{a,b\}/c]&=\frac{\Lambda_2 \psi[\{a,b\},c]}{\Lambda_2+\omega_{ab}+\omega_c}\\
\end{split} 
\end{equation} 

All other GF terms are identical to those in the divergence model without admixture \citep[see eq.\ 1][Appendix, with $\beta=1$]{Lohse:2012}. Using \emph{Mathematica} the set of GF equations is easilsy solved.

We can recover the GF for the original model of discrete split and admixture times which we denote $P[\underline{\omega}]$ from $\psi[\underline{\omega}]$ by noting that $\psi[\underline{\omega}]=\int \Lambda_1\Lambda_2\Lambda_{gf} P[\underline{\omega}]e^{-\underline{\Lambda}.\underline{T}}d\underline{T}$. Thus multiplying $\psi[\underline{\omega}]$ by $(\Lambda_{gf} \Lambda_1 \Lambda_2)^{-1}$ and inverting once for each event with respect to the respective $\Lambda$ parameter gives $P[\underline{\omega}]$. Conditioning on the topology gives:

\begin{small}
\begin{equation}
\begin{split}
P[\omega_2,\omega_3|G_{bc}]&=\frac{e^{-(\tau_1+ T_{gf}) \omega_3}(e^{-\omega_2 \tau_2} (f-1)(3+\omega_3)+e^{-\tau_1 -(1+\omega_3) \tau_2} (e^{\tau_1}(f-1) (2+\omega_2)+f(1-\omega_2+\omega_3)}{(1+\omega_2)(3+\omega_3)(1-\omega_2+\omega_3) }\\
P[\omega_2,\omega_3|G_{ab}]&=\frac{e^{-T_{gf} \omega_3}(e^{-\omega_2 (\tau_1 + \tau_2)} f(3+\omega_3)+e^{-(1+\omega_3)(\tau_1 + \tau_2)} (-f(2+\omega_2)-e^{\tau_1}(f-1) (1-\omega_2+\omega_3)}{(1+\omega_2)(3+\omega_3)(1-\omega_2+\omega_3)}\\
P[\omega_2,\omega_3|G_{ac}]&=\frac{e^{-\tau_1(1+\omega_3)-\tau_2 -\omega_3(\tau_2 + T_{gf})}(-e^{\tau_1}(f-1)+f)}{(1+\omega_2)(3+\omega_3)}\\
\end{split}
\label{E:3}
\end{equation}
\end{small}

The above uses the fact that for each topology, the GF only depends on the intervals between the two coalescence events. For example, for topology $G_{bc}$ we define corresponding dummy variables $\omega_3=\omega_a+\omega_b+\omega_c$ and $\omega_2=\omega_a+\omega_{bc}$. Note also that $\tau_1$ and $\tau_2$ are the times between admixture and divergence events (fig.\ 1A). The corresponding time parameters measuring time from the present are: $T_1 = T_{gf}+ \tau_{1}$ and $T_2 = T_{gf}+\tau_{1}+\tau_{2}$.

Without admixture (i.\ e.\ $f \rightarrow 0$ and $T_{gf} \rightarrow 0$) eq.\ \ref{E:3} reduces to eqs.\ 3 and 4 in \citet{Lohse:2012}. Furthermore, we can find the probability of each topology by setting the $\omega$ terms in eq.\ \ref{E:3} to $0$:

\begin{small}
\begin{equation}
\begin{split}
P_{bc}=&\frac{1}{3}(3-3f+e^{-\tau_1-\tau_2}(2e^{\tau_1}(f-1)+f))\\
P_{ab}=&\frac{1}{3}(e^{-\tau_1-\tau_2}(-e^{\tau_1}(f-1)-2f)+3f)\\
P_{ac}=&\frac{1}{3}e^{-\tau_1-\tau_2}(-e^{\tau_1}(f-1)+f)\\
\end{split}
\label{E:4}
\end{equation}
\end{small}

An alternative derivation of eq.\ \ref{E:4} can be made using discrete-time transition matrices \citep[analogous to][]{Slatkin:2008,LohsePhd:2010}.

The moments of the length of a particular branch can be easily found from the GF by taking derivatives with respect to the dummy variable corresponding to that branch. For example, the expected length of internal branches of genealogies with the two incongruent topologies are $E[t_{ab}]= -\frac{\partial P[\underline{\omega}|G_{ab}]}{\partial \omega_{ab}}\rvert_{\omega_{ab}=0}$ and $E[t_{ac}]= -\frac{\partial P[\underline{\omega}|G_{ac}]}{\partial \omega_{ac}}\rvert_{\omega_{ac}=0}$. Multiplying the above by $\theta/2 = 2 N_e \mu$, gives the expected number of the two incongruent types of shared derived mutations $k_{ab}$ and $k_{ac}$ (i.e.\ $Pr(ABBA)$ and $Pr(BABA)$ in the notation of \citet[][eqs.\ 3 \& 4]{Durand:2011}). 

For simplicity, the model above assumes that both ancestral populations are of the same size. We can relax this assumption (i.e.\ the IUA$_2$ model) by defining a rate of pairwise coalescence in the population between the two population splits $\alpha$ (instead of 1) (see Supporting.nb). The derivation for the AS model is analogous to the above and given in the Supporting.nb. Note that while \citep{Durand:2011} assume symmetric migration across the barrier and an additional time parameter $T$ at which the barrier arises; we consider a slightly simpler model where gene flow across the barrier is unidirectional with rate $M/2$ and follow \citep{Slatkin:2008} in assuming a permanent barrier.

Given an infinite sites mutation model, the probability of a particular mutational configuration in a sequence block $P[\underline{k}_j]$ can be calculated from eq.\ \ref{E:3} by taking successive derivatives \citet{Lohse:2011b,Lohse:2012}. We restrict the computation of exact probabilities to mutational configurations that involve up to a maximum of $k_{m}$ mutations on any one genealogical branch. To speed up the computation, the probabilities of rare configurations with more than $k_{m}$ mutations on one or several branches are combined in the likelihood calculation. Since within each topology class, we distinguish mutations on three branches, there are three classes of such configurations involving more than $k_m$ mutation on one, two or all three branches respectively. Their probabilities are calculated from the GF by subtracting the sum of exact probabilities for all configurations involving up to $k_m$ mutations on a branch (or branches) from the relevant marginal probability. We used a threshold of $k_{m}=3$ per branch throughout. Details are given in \citet{Lohse:2011b,Lohse:2012}. Code for this calculation is implemented in \emph{Mathematica} \citep{Wolfram:2010} (available from the authors on request). The sum of likelihoods across loci is maximized using the inbuilt \emph{Mathematica} function \emph{FindMaximum} (this takes a few minutes on a modern desktop).

\newpage

\footnotesize \bibliography{thesisbiborg}

\begin{table}[h] \small
\caption{The expected information on parameters in the IUA model (for the parameter values assumed by \citet{Durand:2011}, see bottom row in bold) in a sequence block. The second row gives the expected standard deviation of parameter estimates based on 10,000 blocks. Results are shown for two 2kb and 4kb blocks.}
\bigskip

\label{T:1}
\begin{tabular}{lllll|llll}
\toprule
\multicolumn{5}{c}{2kb} & \multicolumn{4}{c}{4kb}\\
\midrule
Parameter					&$T_1$	&$T_2$	&$T_{gf}$ 	&$f$	&$T_1$	&$T_2$	&$T_{gf}$ 	&$f$\\
\midrule		
$E[I]$ 	&0.733	&0.701 	&0.003 	&0.477	&1.27	&1.22 	&0.008 	&0.838\\
$E[SD]$, (10 000 loci)	&0.0117  &0.0119 &0.178 & 0.0145 &0.00886 & 0.0091 & 0.112 & 0.011\\
\citet{Durand:2011} history &\textbf{0.125}	&\textbf{0.15}	&\textbf{0.60}	&\textbf{0.04} 	&&&&\\
\bottomrule
\end{tabular}
\end{table}

\begin{table}[tp] \small
\caption{Maximum likelihood estimates of parameters under the divergence with admixture model allowing ancestral populations to have different effective sizes (IUA$_2$) for two different block schemes. Time parameters are scaled in $2 N_e$ generations; the second row (in bold) gives absolute values, i.\ e.\ effective population sizes in individuals and divergence in KY. 95\% confidence intervals are shown in brackets.}
\bigskip
\bigskip
\label{T:2}
\begin{tabular}{lllllllll}
\toprule
Data		&$\theta$ ($N_1$)	&$\theta$ ($N_2$) &$T_1$			&$T_2$				&$t_{GF}$ 	&$f$ \\
\midrule  
CHB, 4kb		&0.71	&0.97	&0.415	&1.26 		&0.415	&0.067,  (0.054--0.080)\\
			&\textbf{5,950, (5,880--6,030)}	&\textbf{8,080,  (7,700--8,500)}&\textbf{124, (116--131)}	&\textbf{376, (368--383)}		&\textbf{124, (81.6--$T_1$)}	&~\\
CEU, 4kb		&0.17	&0.98	&0.411	&1.27 		&0.411	&0.065, (0.052--0.078) \\
			&\textbf{5,920, (5,840--5,990)}	& \textbf{8,180,  (7,790--8,600)}&\textbf{122, (115--131)}	&\textbf{377, (369--385)}		&\textbf{122, (79.9--$T_1$)}	&~\\
\midrule 

CHB, 8kb		&1.17	&1.86	&0.415	&1.26 		&0.415	&0.059, (0.050--0.068)\\
			&\textbf{4,890, (4,840--4,930)}	&\textbf{7,750,  (7,520--8,000)}&\textbf{137, (132--145)}	&\textbf{401, (395--407)}		&\textbf{137, (112--$T_1$)}	&~\\
CEU, 8kb		&1.17	&1.84	&0.411	&1.27 		&0.411	&0.056, (0.047, 0.064) \\
			&\textbf{4,870, (4,820--4,920)}	&\textbf{7,680, (7,360--8,040)}&\textbf{137, (132--146)}	&\textbf{399, (394--405)}		&\textbf{137, (111--$T_1$)}	&~\\
\bottomrule				
\end{tabular}
\end{table}

\begin{table}[h]
\caption{Expected (top half) and observed (bottom half) frequencies of blocks with a total numbers of mutations $S$ for each of the four topology classes. The expection is derived assuming the model that provided the best fit to the 2kb (N/YRI/CEU) data (Table \ref{T:2}) and closely fits the observed frequencies. Note that 80\% of blocks are topologically unresolved.}
\bigskip
\label{T:3}
\begin{tabular}{lllllllllll}
\toprule
\midrule
$S$ &0&1&2&3&4&5&6&7&8&Total\\
\midrule
(N,(YRI,CEU))& n/a &0.046 & 0.043 & 0.024 & 0.0099 & 0.0036 & 0.0012 & 0.00039 & 0.00012 & 0.13 \\
(YRI,(N,CEU))& n/a & 0.012 & 0.013 & 0.0083 & 0.0040 & 0.0016 & 0.00058 & 0.00019 & 0.000062 & 0.039 \\
(CEU,(N,YRI))&  n/a & 0.0085 & 0.011 & 0.0071 & 0.0035 & 0.0014 & 0.00053 & 0.00018 & 0.000058 & 0.032 \\
Unresolved& 0.36 & 0.28 & 0.12 & 0.037 & 0.0099 & 0.0023 & 0.00050 & 0.00010 & 0.000020 & 0.80\\

\midrule
(N,(YRI,CEU))& n/a &0.052 & 0.046 & 0.023 & 0.0097 & 0.0037 & 0.0013 & 0.00039 & 0.00014 & 0.14 \\
(YRI,(N,CEU))& n/a & 0.015 & 0.015 & 0.0084 & 0.0038 & 0.0016 & 0.00059 & 0.00018 & 0.000052 & 0.045 \\
(CEU,(N,YRI))& n/a& 0.013 & 0.013 & 0.0078 & 0.0036 & 0.0013 & 0.00054 & 0.00020 & 0.000045 & 0.040 \\
Unresolved& 0.36 & 0.26 & 0.11 & 0.036 & 0.011 & 0.0027 & 0.00078 & 0.00024 & 0.000063 & 0.78\\
\bottomrule
\end{tabular}
\end{table}

\newpage

\begin{figure}[h]
\caption{The length distribution of the internal branch ($t_{in}$) and the shorter external branches ($t_{ex}$, i.e.\ those connected to the more recent node in the genealogy) under A) the admixture (IUA) model or B) a model of ancestral structure (AS) (fig.\ 1). Branch length distributions for incongruent genealogies with topology $G_{ab}$ (the frequency of which is increased by admixture or populations structure) are shown as solid lines, those for the alternative incongruent topology $G_{ac}$ as dashed lines. A) is based on the parameters of \citep{Durand:2011} with high admixture ($f=0.2$); the parameters in B) are chosen to give the same expected $D$ value.}
\includegraphics[width=6in]{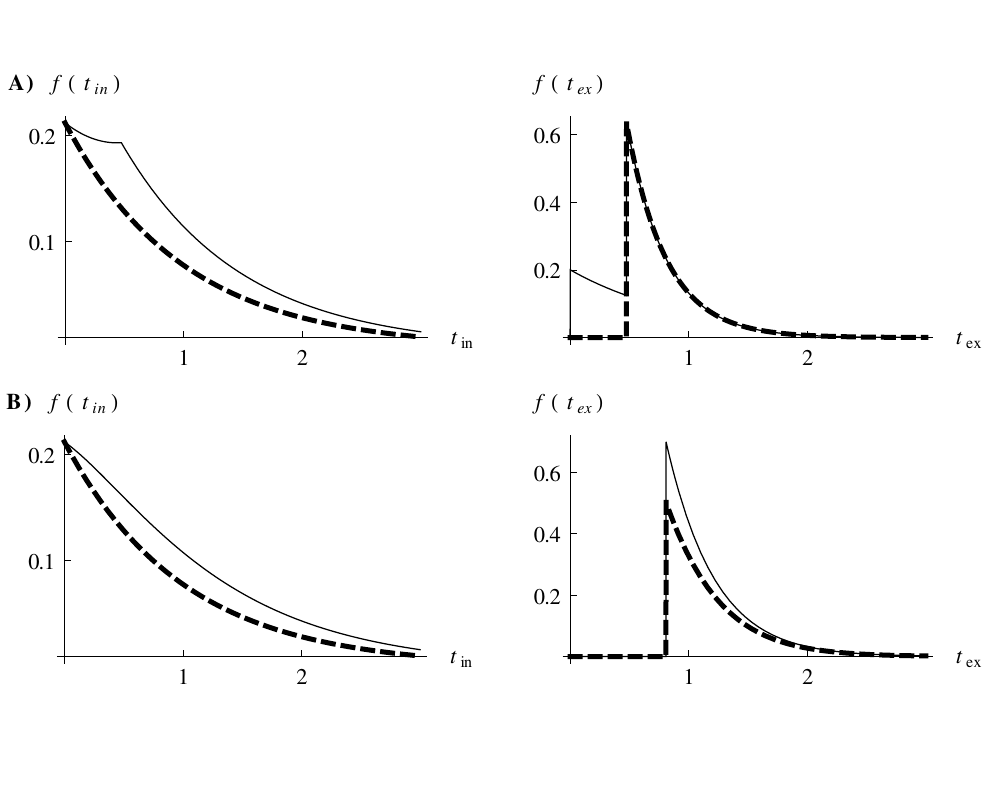}
\label{F:1}
\end{figure}


\begin{figure}[h]
\caption{A) The expected information ($E[\Delta lnL]$) to distinguish the IUA model (\citep{Durand:2011} parameters) from a null model of strict divergence. The dotted line shows the information contained in 10,000 unlinked SNPs. The grey line corresponds to 10,000 blocks each containing a single SNP on average analysed using maximum likelihood. Black, green and red show results for 2kb, 4kb and 8kb blocks respectively. B) The expected standard deviation ($E[SD]$) of $f$ for the likelihood method plotted against block length.}
\includegraphics[totalheight=2.2in]{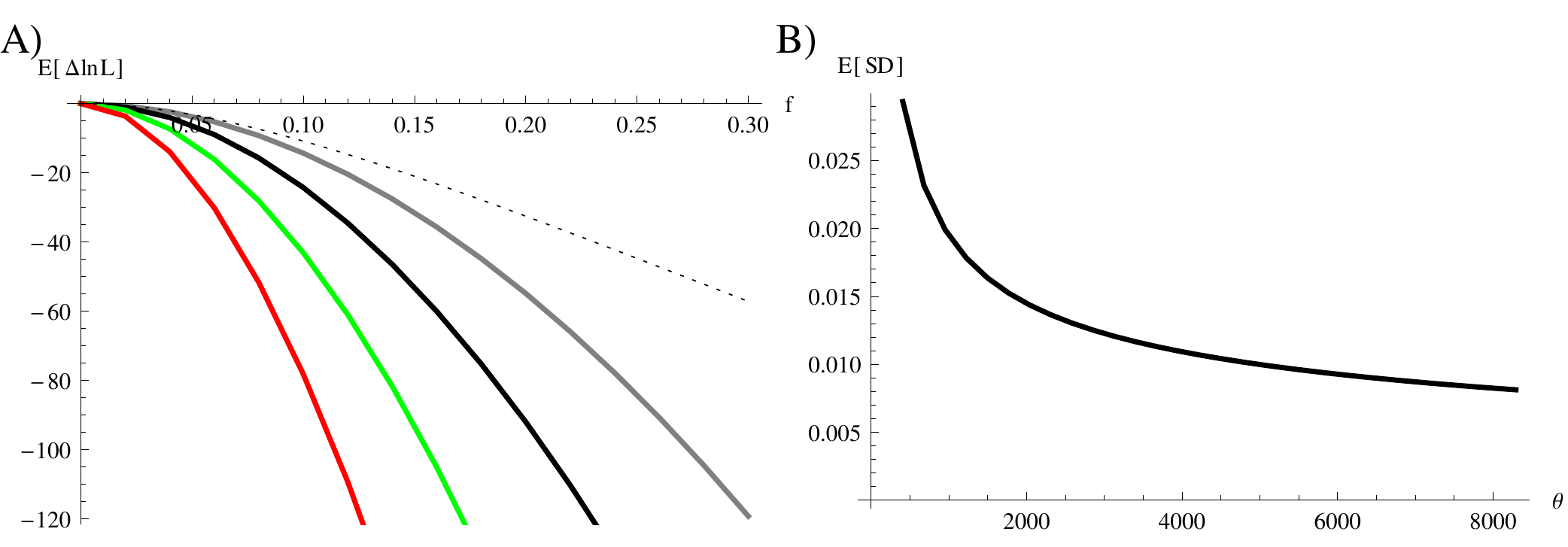}
\label{F:3}
\end{figure}

\begin{figure}[h]
\caption{$\Delta lnL$ plotted against the admixture proportion $f$ (from Neandertals into Eurasians) inferred from the 2 kb (black), 4kb (green) and 8kb data (red) for the CEU (dashed lines) and the CHB (solid) triplets. 95\% confidence intervals are given by the horizontal line.}
\includegraphics[totalheight=3in]{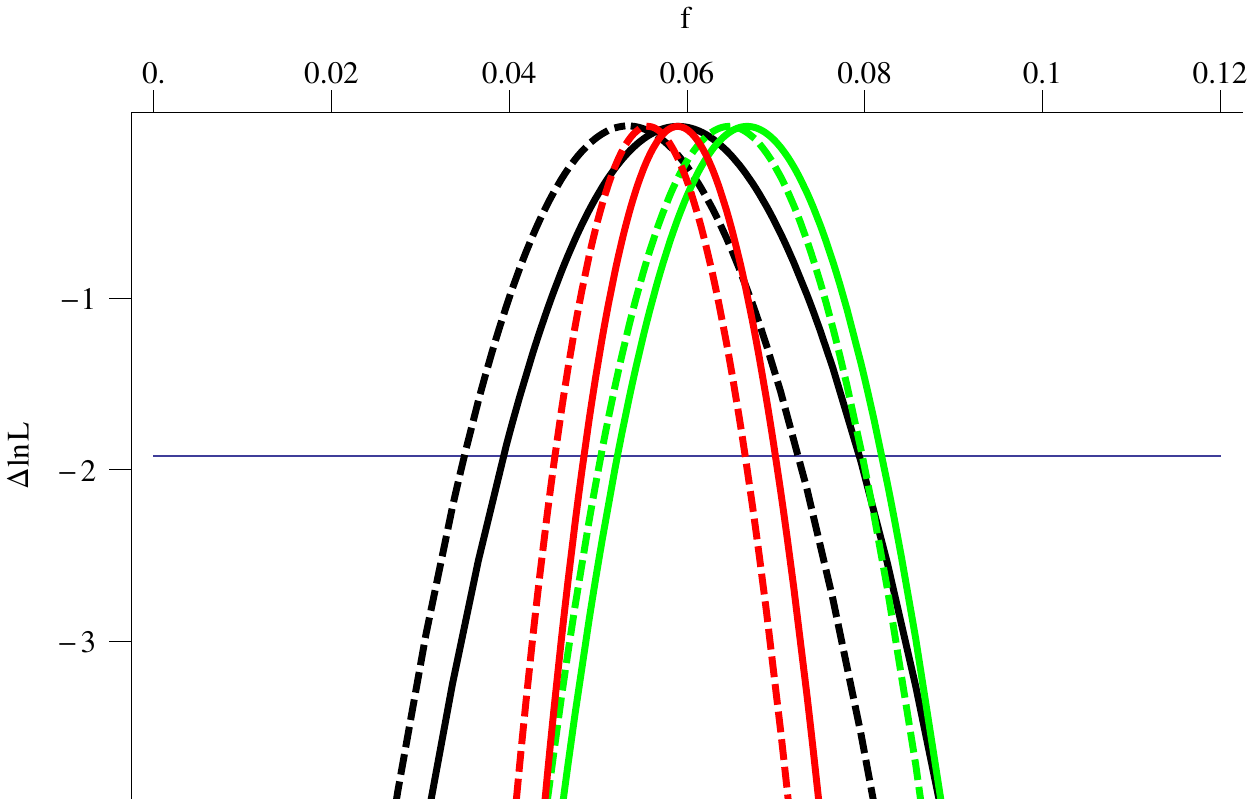}
\label{F:4}
\end{figure}

\begin{figure}[h]
\caption{Expected estimates of parameters from data simulated with recombination (1.3 cM/Mb) plotted against block length. The parameter estimates from the 2, 4 and 8klb analyses of the CEU dataset (assuming no intra-locus recombination) are shown as black dots.}
\includegraphics[totalheight=4in]{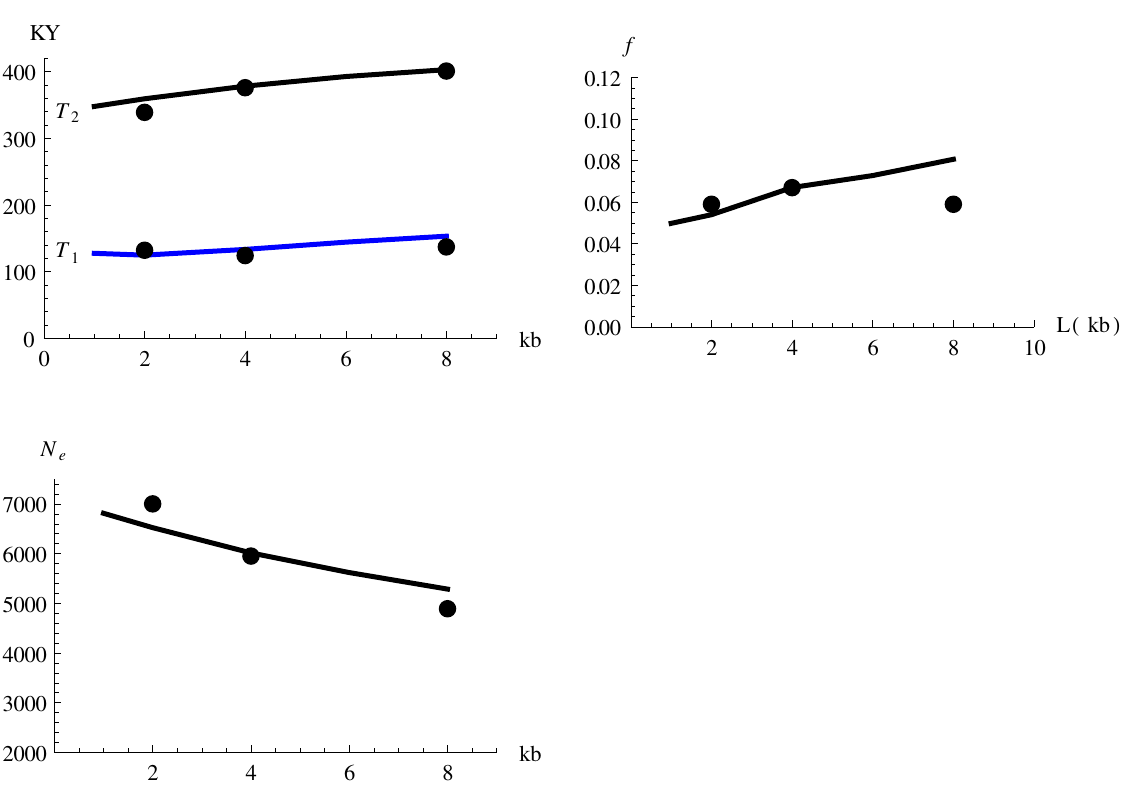}
\label{F:5}
\end{figure}

\begin{figure}[h]
\caption{An example of a genealogy underlying sequence data from three diploid individuals (a, b, and c). Homozygous sites (filled circles) or single heterozygous sites in an individual (white square on the branch leading to b1) present no phasing problem. Similary, multiple heterozygous sites in an individual (green squares) may be phased randomly, assuming an inifinite sites mutation model and at least one diagnostic homozygous site which is in the derived state in this individual only (blue circles). Although phasing such simple hets at random may result in inferring the wrong haplotypes, this cannot introduce biases because the underlying genealogical branches have the same length. This is not the case in the absence of diagnostic sites or for complex hets that are variable in multiple individuals (red circles).}
\includegraphics[totalheight=5in]{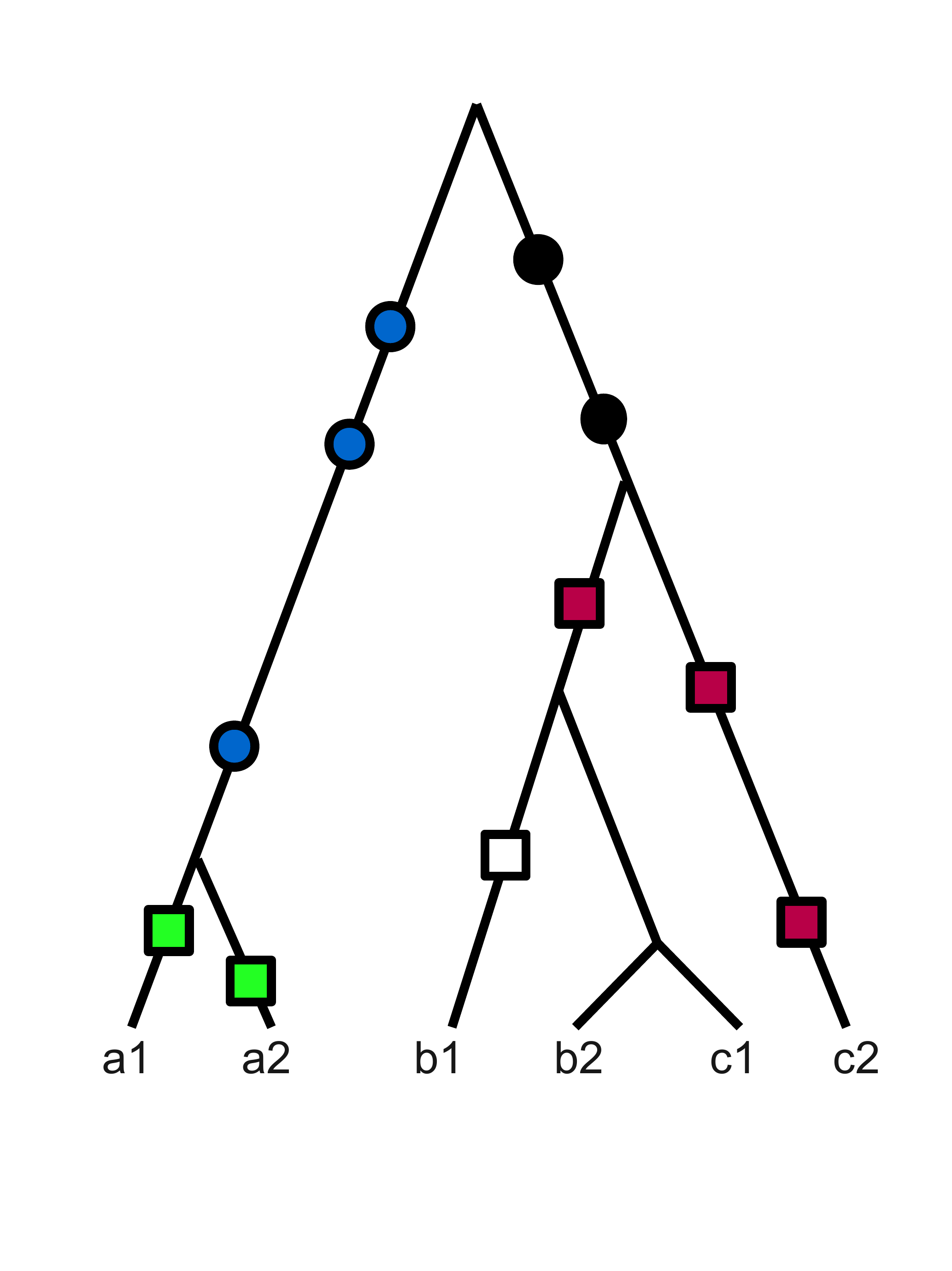}
\label{F:6}
\end{figure}